# THERMODYNAMICS OF MIXTURES CONTAINING ALKOXYETHANOLS. XXVIII. LIQUID-LIQUID EQUILIBRIA FOR 2-PHENOXYETHANOL + SELECTED ALKANES

VÍCTOR ALONSO, MARIO GARCÍA, JUAN ANTONIO GONZÁLEZ*, ISAÍAS GARCÍA DE LA FUENTE AND JOSÉ CARLOS COBOS

**G.E.T.E.F., Grupo Especializado en Termodinámica de Equilibrio entre Fases**, Departamento de Física Aplicada, Facultad de Ciencias, Universidad de Valladolid, E-47071, Valladolid (Spain).

*corresponding author
E-mail: jagl@termo.uva.es

**ABSTRACT**

The coexistence curves of the liquid-liquid equilibria (LLE) for systems of 2-phenoxyethanol (2PhEE) with heptane, octane, cyclohexane, methylcyclohexane or ethylcyclohexane have been determined by the method of the critical opalescence using a laser scattering technique. All the curves show an upper critical solution temperature (UCST), have a rather horizontal top and their symmetry depends on the relative size of the mixture compounds. UCST values are higher for systems with linear alkanes than for solutions including cyclic alkanes. For these mixtures, the UCST increases with the size of the alkyl group attached to the cyclic part of the molecule. It is shown that interactions between alkoxyethanol molecules are stronger when the hydroxyether contains an aromatic group. Data are used to determine the critical exponent for the order parameter mole fraction. Values obtained are consistent with those provided by the Ising model or by the renormalization group theory.

KEYWORDS: LLE; 2-phenoxyethanol; alkanes/ UCST/ dipolar interactions

## 1. INTRODUCTION

Alkoxyethanols, $CH_3\text{-}(CH_2)_n\text{-}O\text{-}(CH_2\text{-}CH_2O)_m\text{-}CH_2\text{-}CH_2OH$, are a very interesting class of substances from a practical point of view, as oxygenated compounds are increasingly used as additives to gasoline due their octane-enhancing and pollution-reducing properties [1,2]. In addition, hydroxyethers are non-ionic amphiphile molecules, very effective as surfactants with a large number of applications [3,4]. On the other hand, alkoxyethanols are also of interest due to the presence of the O and OH groups in the same molecule, which allows self-association via inter- and intramolecular hydrogen bonds. Different spectroscopic techniques have been used to investigate the existence of intramolecular hydrogen bonds [5-8]. They are present at all conditions, even in vapour phase, whereas intermolecular H-bonds appear at higher concentrations of the cellosolve in liquid state. Thus, $CH_3\text{-}(CH_2)_n\text{-}O\text{-}(CH_2)_p\text{-}OH$ molecules can form 5-, 6- and 7-membered rings for p = 2, 3 and 4, respectively. The formation of intramolecular H-bonds is more favourable when the molecules are in gauche conformations. The stability of these conformations decreases with increasing p, resulting in a smaller number of rotational conformations favourable for intramolecular H-bonds. On the other hand, the increasing ring size of the intramolecularly H-bonded monomers results in greater restriction to internal rotation and is responsible for increasing the entropy changes involved in the interconversion of the mentioned species into free monomers. Alkoxyethanols with two ether groups and m = 1 form 5-membered rings similar to those previously cited, but can also form 8-membered rings of quite different properties. This can be ascribed to the replacement of a methylene group by a less bulky oxygen atom which increases the stability of the ring system by decreasing the ring strain caused by crowding repulsions [6].

The formation of the intramolecular H-bonds leads to enhanced dipole-dipole interactions in solutions containing alkoxyethanols and alkanes in comparison to those present in mixtures with homomorphic alkanols [9]. The following evidences support this statement. a) The Trouton's constant of hydroxyethers, 99.58 J $mol^{-1}$ $K^{-1}$ [10], is closer to the value of non-self-associated compounds, 92.05 J $mol^{-1}$ $K^{-1}$ [10], than the value for 1-alkanols, 110.88 J $mol^{-1}$ $K^{-1}$ [10]. b) For a given alkane, the upper critical solution temperature (UCST) of the system including an alkoxyethanol is higher than that of the solution with the homomorphic 1-alkanol [9,11-14]. c) $H_{m,1}^{E,\infty}$ (2-alkoxyethanol(1) + alkane(2)) < $H_{m,1}^{E,\infty}$ (1-alkanol(1) + alkane(2)) [9,15], where $H_{m,i}^{E,\infty}$ is the excess partial enthalpy at infinite dilution of component i. For example, at 298.15 K, $H_1^{E,\infty}$ (2-methoxyethanol + heptane) = 17.6 kJ $mol^{-1}$ [15]. For 1-alkanol + alkane mixtures, this magnitude, independent of the alkane [16,17] is, at 303.15 K, 23.4 kJ $mol^{-1}$ for 1-hexanol [18], and 21.8 for 1-octanol [19]. This has been interpreted assuming that the hydroxyether molecules become free via the destruction of the intermolecular H-bonds on addition of the solvent molecules, and that each alkoxyethanol molecule forms an intramolecular H-bond between the O and the OH group of the same molecule and thus stabilizes itself [15]. d) The molar excess enthalpies, $H_\mathrm{m}^\mathrm{E}$, of alkoxyethanol + alkane mixtures are higher than those of homomorphic 1-alkanol + alkane systems and the $H_\mathrm{m}^\mathrm{E}$ curves for systems with alkoxyethanols are more symmetrical than those of alcoholic solutions, skewed towards low concentration of the self-associated compound [9].

In previous works, we have reported data on $H_\mathrm{m}^\mathrm{E}$, $C_{\mathrm{p,m}}^\mathrm{E}$ (molar excess heat capacities at constant pressure), $V_\mathrm{m}^\mathrm{E}$ (excess molar volumes), compressibilities and speeds of sound, vapour-liquid equilibria and LLE for binary mixtures of alkoxyethanols with different organic solvents (alkanes, 1-butanol, dibutylether, hydroxyethers) [11-14,20-23], and we have investigated these complex systems in terms of different theories: DISQUAC [9,24], ERAS [24,25], Flory [26-28], or the Kirkwood-Buff formalism [28]. As a continuation, we report here LLE data for 2-phenoxyethanol + alkane mixtures. This is useful to gain insight into the influence of the aromatic ring on the thermodynamic properties of alkoxyethanols.

## 2. EXPERIMENTAL

### 2.1 Materials

Heptane (142-82-5, puriss p.a. ≥ 0.995); octane (111-65-9, puriss p.a. ≥ 0.99); cyclohexane (110-82-7, puriss p.a. ≥ 0.995); methylcyclohexane (108-87-2, puriss p.a. ≥ 0.98) were from Fluka; ethylcyclohexane (167-891-7, puriss p.a. ≥ 0.99) and 2-phenoxyethanol (122-99) were from Aldrich (purities expressed in mass fraction). Prior to the measurements, the chemicals were stored over molecular sieves (UOP 3Å from Fluka). All these chemicals were used without other further treatment. The densities ρ at 298.15 K and atmospheric pressure were in good agreement with literature values (Table 1). Regarding to 2PhEE, no density data have been encountered for comparison. For a better characterization of this compound, we have determined its purity by GC analysis (≥ 0.995) and the density and refractive index, $n_\mathrm{D}$, at different temperatures. At (293.15 and 303.15) K, ρ/g·cm$^3$ = 1.10692 and 1.09916, while $n_\mathrm{D}$ = 1.53907, 1.53702 and 1.53501 at (293.15, 298.15 and 303.15) K, respectively. The water contents, determined by the Karl-Fischer method, were as follows (in mass fraction): 5·10$^{-6}$ for heptane, cyclohexane and methylcyclohexane, 26·10$^{-6}$ for octane, 46·10$^{-6}$ for ethylcyclohexane and 19·10$^{-6}$ for 2-phenoxyethanol.

### 2.2 Apparatus and Procedure

Mixtures were prepared by mass, with a weighing accuracy to ± 0.00001 g, in Pyrex tubes of 1 cm i.d. and about 4 cm length, which then were immediately sealed by capping at atmospheric pressure and room temperature. Conversion to molar quantities was based on the relative atomic mass table of 1996 issued by IUPAC in 1996 [29].

The coexistence curves of the binary mixtures were determined by the method of the critical opalescence. The samples in the sealed Pyrex tubes are placed in a thermostat bath a few hundredths of degree above the expected temperature. A red He-Ne laser (wavelength of 635 nm) is placed on one side of the equilibrium cell so that the light beam passing through the solution is focused on a photodiode (with its maximum response around 700 nm). When decreasing slowly the temperature (1.2 K h$^{-1}$), the growth of small drops of the dispersed liquid phase causes a diffusion of light during the transition. This results in a voltage variation in the photodiode output, measured by a digital multimeter model Agilent 34410A connected to a PC, which allows simultaneous accurate measurements of the transition temperatures. The temperature was measured with a precision of ± 0.001 K and estimated accuracy of ± 0.05 K by a Pt-1000 resistance connected to a multimeter model Philips PM2353, in such way that the resistance variations upon cooling are also registered in a PC. The thermometer was calibrated on the basis of the ITS-90 scale of temperature using the triple point of the water, and the fusion point of Ga. The separation temperatures were reproducible to ± 0.05 K for temperatures near the upper critical solution temperature. The precision of the equilibrium composition is expected to be better than 0.0005 in mole fraction. The weighing technique gives a precision better than 0.0001 in mole fraction, but this is reduced slightly due to partial evaporation of the more volatile component to the free volume of the ampoule (≈1.4 cm$^3$).

3.    RESULTS

Table 2 lists the direct experimental results of the liquid-liquid equilibrium temperatures $T$ vs. the mole fraction of 2PhEE, $x_1$, for the investigated mixtures (Figs. 1,2).

All the systems show an UCST. The LLE coexistence curves have a flat maximum, and their symmetry depends on the relative size of the mixture compounds (Figs. 1,2). Thus, an increase of the molar volume of the alkane leads to LLE curves which become progressively skewed towards increasing $x_1$ values (Figs. 1,2). As consequence, the representation of the equilibrium temperatures vs. $x_1$ or vs. $\Phi_1$, the volume fraction ($\Phi_1 = x_1V_i / (x_1V_1 + x_2V_2)$, $V_i$, molar volume of component i) differ for the octane system (Fig. 1).

The coordinates of the critical points, $x_{1c}$ and $T_c$ (Table 3), were obtained by reducing the experimental data with the equation [30,31]

$$T / K = T_c / K + k|y - y_c|^m \qquad (1)$$

where

$$y = \frac{\alpha \, x_1}{1 + x_1(\alpha - 1)} \quad (2)$$

$$y_c = \frac{\alpha \, x_{1c}}{1 + x_{1c}(\alpha - 1)} \quad (3)$$

In eqs 1 to 3, $m$, $k$, $\alpha$, $T_c$ and $x_{1c}$ are the coefficients to be fitted to the experimental results. When $\alpha = 1$, eq (1) is similar to the well-known equation [32-35]:

$$\Delta \lambda = B \tau^{\beta} \quad (4)$$

where $\Delta \lambda_1 = \lambda_1' - \lambda_2''$ is the so-called order parameter, which can be any density variable in the conjugate phase [35] (mole fraction, volume fraction; in our case $\lambda_1 = x_1$), $\tau$ is the reduced temperature $(T_c - T)/T_c$ and $\beta$ a critical exponent corresponding to this order parameter. The $\beta$ value depends on the theory applied to its determination [33-37]. More details are given below.

The fitting was performed using the Marquardt algorithm [38] with all the points weighted equally. Results are collected in Table 3. Also listed is the standard deviation defined by:

$$(\sigma(T)/K) = \left[ \sum (T_{\exp,i} - T_{\mathrm{calc},i})^2 / (N - n) \right]^{1/2} \quad (5)$$

where $N$ and $n$ stand for the number of data points and the number of fitted parameters, respectively. We note that eq 1 fits well the experimental data.

## 4.  DISCUSSION

We note that the UCSTs of mixtures involving linear alkanes are higher than those of systems with cyclic alkanes (Table 3). The same trend is observed in solutions containing other alkoxyethanols. Thus, for 2-methoxyethanol (2ME) mixtures, UCST(cyclohexane) = 294.58 K [11] < UCST(hexane) = 311.19 K [11] and UCST(methylcyclohexane) = 297.34 K [13] < UCST(heptane) = 319.74 K [13]. This reveals that cyclic alkanes break more easily interactions between alkoxyethanol molecules than $n$-alkanes. Consequently, UCST of 2PhEE solutions increases in the sequence: cyclohexane < methylcyclohexane < ethylcyclohexane (Table 3, Fig 2), as the size of the alkyl group attached to the cyclic part of the molecule increases in the same order. For a similar reason, UCST(heptane) < UCST(octane) (Table 3). The increase of UCST with the chain length of the *n*-alkane is also encountered in mixtures with 2ME, 2-ethoxyethanol (2EE), or 2-(2-ethoxyethoxy)ethanol (22EEE) [11,12,14].

It is interesting to compare UCST of binary systems containing different alkoxyethanols and the same alkane. Thus, for systems with heptane, UCST increases in the order: 261.15 K (2EE) [39] < 287.97 K (22EEE) [14] < 319.74 K (2ME) [13]] < 367.57 K (2PhEE) (this work). This indicates that interactions between like molecules are enhanced by the presence of the aromatic group in 2-phenoxyethanol.

As previously mentioned, the $\beta$ value in eq. (4) depends on the theory applied for its determination. The renormalization group approach yields 0.325, while the Ising model gives 0.312. The classical model gives 0.5 [32-37]. Application of eq. (1) is rather difficult because it is needed to know the composition in conjugate phases to determine the order parameter. The experimental method followed in this work suffers from the drawn-back that the equilibrium temperatures of the two coexisting compositions are not determined simultaneously. Here, we have used a modification of eq. (1) to estimate $\beta$, which is obtained taking $\alpha = 1$ [32,40,41]:

$$T/K = T_c/K + k'|x_1 - x_{1c}|^{1/\beta} \qquad (6)$$

When applying this equation, firstly, it is necessary to determine its validity range. This was done by analyzing the data for various choices of range [42]. Thus, eq. (6) was used for temperature ranges ($T_c - T$)/K= 1.96, 1.63, 1.57, 1.94 and 1.80 for the systems with heptane, octane, cyclohexane, methylcyclohexane, or ethylcyclohexane, respectively. The fitting of the four parameters in eq. (6), $T_c$, $x_{1c}$, $k'$, and $1/\beta$ was developed as previously. The standard deviations were (0.05, 0.05, 0.05, 0.02 and 0.02) K for the mixtures in the same order as cited above. The $\beta$ values calculated were: 0.307 (heptane); 0.299 (octane); 0.362 (cyclohexane), 0.314 (methylcyclohexane) and 0.304 (ethylcyclohexane). Similar results have been obtained for systems such as: methanol + heptane, or + carbon disulphide, acetic anhydride + heptane, + cyclohexane, or + carbon disulphide [40,41]. It is interesting to note that no improvement is reached when replacing the mole fraction by the volume fraction, which is probably due to the similar molar volumes of the mixtures components.

### 5. CONCLUSIONS

LLE coexistence curves were determined for mixtures of 2PhEE with heptane, octane, cyclohexane, methylcyclohexane or ethylcyclohexane. UCST values are higher for systems with linear alkanes than for solutions including cyclic alkanes. For the latter mixtures, the UCST increases with the size of the alkyl group attached. Interactions between alkoxyethanol molecules are stronger when the hydroxyether contains an aromatic group.

## 7. LIST OF SYMBOLS

| | |
|---|---|
| $T$ | equilibrium temperature |
| $k, m$ | parameters in eq. (1) |
| $x$ | mole fraction in liquid phase |

*Greek letters*

| | |
|---|---|
| $\alpha$ | parameter in eq.(1) |
| $\beta$ | critical exponent for the order parameter (mole fraction) |
| $\rho$ | density |
| $\sigma$ | relative standard deviation (eq. 5) |
| $\tau$ | reduced temperature |

*Subscripts*

| | |
|---|---|
| c | critical point |

**ACKNOWLEDGEMENTS**

The authors gratefully acknowledge the financial support received from the Consejería de Educación y Cultura of Junta de Castilla y León, under Project VA052A09 and from the Ministerio de Educación y Ciencia, under the Project FIS2010-16957. V.A. also gratefully acknowledges the grant received from the Junta de Castilla y León. Authors acknowledge technical assistance from Felipe Sanz.

TABLE 1

Comparison of experimental densities $\rho$ with literature for pure liquids at $T = 298.15$ K

| Compound | $\rho$ / (g·cm$^{-3}$) | |
| --- | --- | --- |
| | Experimental | Literature [43] |
| 2-Phenoxyethanol | 1.10341 | |
| Heptane | 0.67971 | 0.67946 |
| Octane | 0.69876 | 0.69862 |
| Cyclohexane | 0.77395 | 0.77389 |
| Methylcyclohexane | 0.76478 | 0.76506 |
| Ethylcyclohexane | 0.78402 | 0.7839 |

TABLE 2

Experimental liquid-liquid equilibrium temperatures for 2-phenoxyethanol(1) + alkane(2) mixtures.

| $x_1$ | $T$/K | $x_1$ | $T$/K |
|---|---|---|---|
| 2-phenoxyethanol(1) + heptane(2) | | | |
| 0.2089 | 361.59 | 0.4602 | 367.56 |
| 0.2389 | 363.41 | 0.4752 | 367.54 |
| 0.2618 | 364.76 | 0.4976 | 367.60 |
| 0.2810 | 365.64 | 0.5112 | 367.45 |
| 0.3007 | 366.37 | 0.5250 | 367.35 |
| 0.3379 | 367.02 | 0.5466 | 367.28 |
| 0.3741 | 367.38 | 0.5926 | 366.41 |
| 0.3954 | 367.52 | 0.6273 | 365.13 |
| 0.4232 | 367.54 | 0.6537 | 363.48 |
| 0.4297 | 367.53 | 0.6858 | 361.45 |
| 0.4511 | 367.59 | | |
| 2-phenoxyethanol(1) + octane(2) | | | |
| 0.2774 | 365.92 | 0.4740 | 369.25 |
| 0.2914 | 366.52 | 0.4998 | 369.25 |
| 0.2992 | 366.86 | 0.5221 | 369.16 |
| 0.3196 | 367.62 | 0.5446 | 369.17 |
| 0.3382 | 368.19 | 0.5700 | 368.98 |
| 0.3561 | 368.46 | 0.5914 | 368.80 |
| 0.3719 | 368.84 | 0.6195 | 368.26 |
| 0.3958 | 368.95 | 0.6406 | 367.64 |
| 0.4208 | 369.12 | 0.6687 | 366.74 |
| 0.4448 | 369.18 | 0.6894 | 365.69 |
| 2-phenoxyethanol(1) + cyclohexane(2) | | | |
| 0.1166 | 310.55 | 0.3837 | 314.83 |
| 0.1264 | 311.48 | 0.4145 | 314.70 |
| 0.1562 | 312.96 | 0.4336 | 314.40 |
| 0.1851 | 313.90 | 0.4465 | 314.33 |
| 0.2063 | 314.36 | 0.4642 | 314.04 |
| 0.2264 | 314.63 | 0.4788 | 313.81 |
| 0.2555 | 314.88 | 0.4959 | 313.36 |
| 0.2855 | 314.93 | 0.5134 | 312.83 |

TABLE 2 (continued)

| | | | |
|---|---|---|---|
| 0.3246 | 314.93 | 0.5370 | 312.03 |
| 0.3546 | 314.92 | 0.5899 | 309.22 |
| 0.3701 | 314.86 | | |
| 2-phenoxyethanol(1) + methylcyclohexane(2) | | | |
| 0.1527 | 320.99 | 0.4077 | 325.53 |
| 0.1821 | 322.82 | 0.4339 | 325.49 |
| 0.2155 | 324.17 | 0.4586 | 325.39 |
| 0.2404 | 324.77 | 0.4807 | 325.20 |
| 0.2744 | 325.22 | 0.5013 | 324.91 |
| 0.2994 | 325.47 | 0.5278 | 324.44 |
| 0.3176 | 325.51 | 0.5539 | 323.65 |
| 0.3460 | 325.54 | 0.5845 | 322.38 |
| 0.3771 | 325.59 | 0.6144 | 321.13 |
| 2-phenoxyethanol(1) + ethylcyclohexane(2) | | | |
| 0.1990 | 327.34 | 0.4255 | 330.37 |
| 0.2106 | 327.79 | 0.4491 | 330.31 |
| 0.2246 | 328.36 | 0.4756 | 330.32 |
| 0.2499 | 329.09 | 0.4994 | 330.19 |
| 0.2748 | 329.63 | 0.5238 | 330.05 |
| 0.2971 | 329.97 | 0.5497 | 329.70 |
| 0.3252 | 330.20 | 0.5710 | 329.36 |
| 0.3493 | 330.28 | 0.6005 | 328.57 |
| 0.3746 | 330.35 | 0.6233 | 327.79 |
| 0.4027 | 330.36 | 0.6350 | 327.19 |

TABLE 3

Coefficients in eq. (1) for the fitting of the ($x_1$, $T$) pairs given in Table 2 for 2-phenoxyethanol(1) + alkane(2) mixtures; $\sigma$ is the standard deviation defined by eq 5.

| $N^a$ | $m$ | $K$ | $\alpha$ | $T_c$/K | $x_{1c}$ | $\sigma$/K |
|---|---|---|---|---|---|---|
| | | 2-phenoxyethanol(1) + heptane(2) | | | | |
| 21 | 3.16 | $-579.$ | 0.954 | 367.6 | 0.449 | 0.09 |
| | | 2-phenoxyethanol(1) + octane(2) | | | | |
| 20 | 3.17 | $-509$ | 1.005 | 369.2 | 0.481 | 0.05 |
| | | 2-phenoxyethanol(1) + cyclohexane(2) | | | | |
| 21 | 3.34 | -447 | 1.513 | 314.9 | 0.319 | 0.06 |
| | | 2-phenoxyethanol(1) + methylcyclohexane(2) | | | | |
| 18 | 3.22 | $-436$ | 1.310 | 325.6 | 0.369 | 0.06 |
| | | 2-phenoxyethanol(1) + ethylcyclohexane(2) | | | | |
| 20 | 3.30 | $-462$ | 1.006 | 330.3 | 0.415 | 0.03 |

[a] number of experimental data points.

**CAPTION TO FIGURES**

**FIG. 1**  LLE for the 2-phenoxyethanol(1) + octane(2) mixture. Points, experimental results (this work): $T$ vs. $x_1$ (●) or vs. $\Phi_1$ (O). Solid lines, results from the fitting equation (1) using the ($x_1$, $T$) pairs.

**FIG. 2**  LLE for the 2-phenoxyethanol(1) + alkane(2) mixtures. Point, experimental results (this work). Full symbols, $T$ vs. $x_1$: (●), cyclohexane, (■), methylcyclohexane; (▲), ethylcyclohexane; (O), $T$ vs. $\Phi_1$ for the cyclohexane system. Solid lines, results from the fitting equation (1) using the ($x_1$, $T$) pairs.

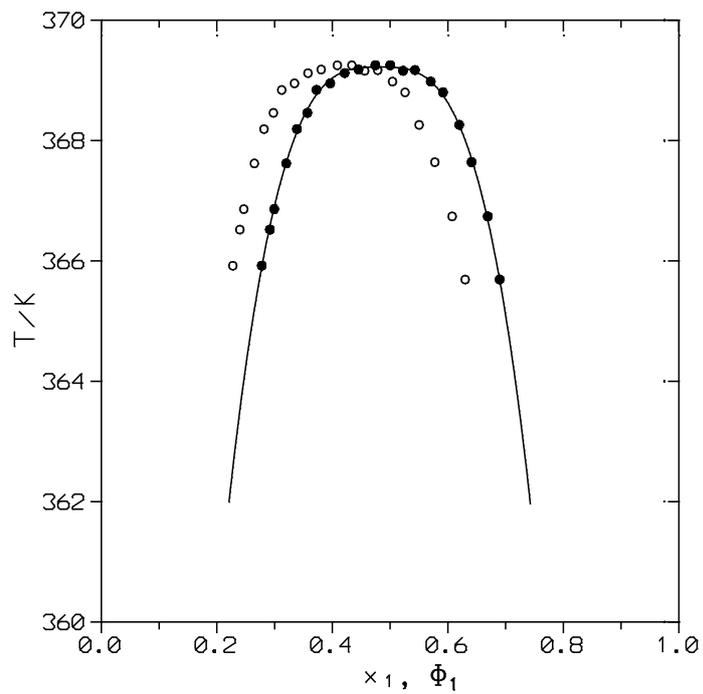

FIG. 1

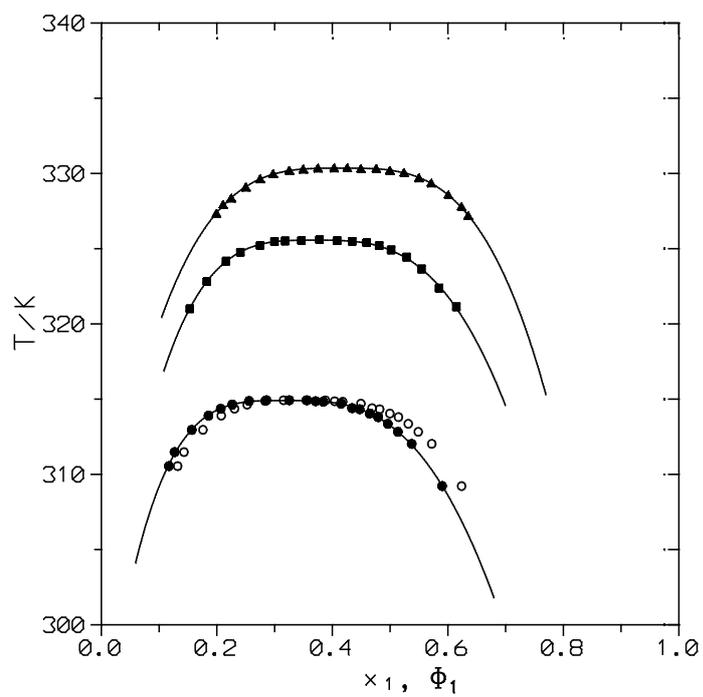

FIG. 2